\documentstyle[12pt, aaspp]{article}
\begin{document}
 
\def\today{\number\year\space \ifcase\month\or  January\or February\or
        March\or April\or May\or June\or July\or August\or
September\or
        October\or November\or December\fi\space \number\day}
\def\fraction#1/#2{\leavevmode\kern.1em
 \raise.5ex\hbox{\the\scriptfont0 #1}\kern-.1em
 /\kern-.15em\lower.25ex\hbox{\the\scriptfont0 #2}}
\def\spose#1{\hbox to 0pt{#1\hss}}
\def\simlt{\mathrel{\spose{\lower 3pt\hbox{$\mathchar''218$}}
     \raise 2.0pt\hbox{$\mathchar''13C$}}}
\def\simgt{\mathrel{\spose{\lower 3pt\hbox{$\mathchar''218$}}
     \raise 2.0pt\hbox{$\mathchar''13E$}}}
\def\etal{et al. }

\title{On the Snow Line in Dusty Protoplanetary Disks}
\author{D. D. Sasselov \& M. Lecar}
\affil{Harvard-Smithsonian Center for Astrophysics, 60 Garden St., Cambridge MA 02138}

\begin{abstract}

Hayashi (1981) prescribed a `minimum-mass solar nebula' which contained
just enough material to make the planets. This prescription, which has
been widely used in constructing scenarios for planet formation, 
proposed that ice will condense when the temperature falls below 
170~K (the ``snow line"). In Hayashi's model that occured at 2.7$AU$.
It is usually assumed that the cores of the giant planets, e.g., Jupiter,
form beyond the snow line.

The snow line, in Hayashi's model, is where the temperature of a black
body that absorbed direct sunlight and re-radiated as much as it absorbed,
would be 170~K. Since Hayashi, there have been a series of more
detailed models of the absorption by dust of the stellar radiation, and
of accretional heating, which alter the location of the snow line. We have
attempted a ``self-consistent" model of a T Tauri disk in the sense that
we used dust properties and calculated surface temperatures that matched
observed disks. We then calculated the midplane temperature for those
disks, with no accretional heating or with small 
($\leq 10^{-8} M_{\odot} {\rm yr}^{-1}$)
accretion rates. (Larger accretion rates can push the snow line out to beyond
4$AU$ but do not match the observed disks). Our models bring the snow line
in to the neighbourhood of 1 AU; not far enough to explain the close
planetary companions to other stars, but much closer than in recent starting 
lines for orbit migration scenarios.

\end{abstract}

\section{Introduction}
Planetary systems are formed from rotating protoplanetary disks, which are
the evolved phase of circumstellar disks produced during the collapse of
a protostellar cloud with some angular momentum.

A standard model of such a protoplanetary disk, is that
of a steady-state disk in vertical hydrostatic equilibrium, with gas and 
dust fully mixed and thermally coupled (Kenyon \& Hartmann 1987). Such
a disk is flared, not flat, but still geometrically thin in the sense defined
by Pringle (1981). The disk intercepts a significant
amount of radiation from the central star, but other heating sources (e.g.
viscous dissipation) can be more important. If dissipation due to mass
accretion is high, it becomes the main source of heating. Such are the
protoplanetary disks envisioned by Boss (1996, 1998), which have relatively
hot (midplane temperature $T_{\rm m}>$ 1200~K) inner regions due to mass
accretion rates of $\sim 10^{-6}$ to $10^{-5} M_{\odot} {\rm yr}^{-1}$.
However, typical T Tauri disks of age $\sim$1~Myr
seem to have much lower mass accretion rates
($\leq 10^{-8} M_{\odot} {\rm yr}^{-1}$) with all other characteristics
of protoplanetary disks (Hartmann \etal 1998, D'Alessio \etal 1998).
For disks of such low accretion rates stellar irradiation becomes increasingly
the dominant source of heating, to the limit of a passive disk modeled by
Chiang \&
Goldreich (1997). In our paper we will confine our attention to the latter
case, without entering the discussion about mass accretion rates.

The optical depth in the midplane of the disk is very high in the radial
direction, hence the temperature structure there is governed by the
reprocessed irradiation of the disk surface. This is the case of a passive
disk (no accretion). At some point along the radial direction the temperature in the
midplane would drop below the ice sublimation level -- Hayashi (1981)
called it the "snow line". 

In this paper we revisit the calculation of the "snow line" for a protosolar
protoplanetry disk, given its special role in the process of planet formation.
We pay particular emphasis on the issues involved in treating the radiative
transfer and the dust properties.

\section{The Model} 
Our model is that of a star surrounded by a flared disk. In this paper, we
have chosen two examples -- a passive disk and a disk with a 
$10^{-8} M_{\odot} {\rm yr}^{-1}$ accretion rate.  Both have the same 
central star of effective
temperature $T_*$= 4000~K, mass $M_*$= 0.5~$M_{\odot}$, and radius
$R_*$= 2.5~$R_{\odot}$. Thus they correspond to the examples used by
Chiang \& Goldreich (1997) and D'Alessio \etal (1998), respectively.
Our disk has a surface gas mass density
$\Sigma$= $r^{-3/2}{\Sigma}_0$, with $r$ in AU and 
${\Sigma}_0$= 10$^3$g~cm$^{-2}$ for our standard minimum-mass solar
nebula model; we varied ${\Sigma}_0$ between 10$^2$ and 10$^4$g~cm$^{-2}$
to explore the effect of disk mass on the results.

The emergent
spectrum of the star is calculated with a stellar model atmosphere code
with Kurucz (1992) line lists and opacities. The disk intercepts the
stellar radiation $F_{irr}(r)$ at a small grazing angle $\phi(r)$ (defined in \S 5).
The emergent stellar spectrum
is input into a code which solves the continuum radiative transfer problem
for a dusty envelope. The solution is a general spherical geometry solution
with a modification of the equations to a section corresponding to a flared
disk (see Menshchikov \& Henning (1997) for a similar approach). 
In that sense, the radiative transfer is solved essentially in 1D (vertically),
as opposed to a full-scale consistent 2D case. The appeal of our approach
is in the detailed radiative transfer allowed by the 1D scheme.

The continuum radiative transfer problem for a dusty envelope is solved
with the method developed by Ivezi\'c \& Elitzur (1997).
The scale invariance applied in the method is
practically useful when the absorption coefficient is independent of
intensity, which is the case of dust continuum radiation. The energy
density is computed for all radial grid points through matrix inversion,
i.e. a direct solution to the full scattering problem. This is both very
fast and accurate at high optical depths.
Note that in our calculations (at $r \geq 0.1$~AU) the temperatures never
exceed 1500-1800K in the disk and we do not consider dust sublimation;
the dust is present at all times and is the dominant opacity source. As in
the detailed work by Calvet \etal (1991) and more recently by D'Alessio
\etal (1998), the frequency ranges of scattering and emission can be treated
separately.

For the disk with mass accretion, the energy rate per unit volume
generated locally by viscous stress is given by
$2.25 \alpha P(z){\Omega}(r)$, where the turbulent viscosity coefficient
is ${\nu}= {\alpha}c_{\rm s}^2\Omega^{-1}$, $\Omega$ is the Keplerian angular
velocity, ${c_{\rm s}}^2=P{\rho}^{-1}$ is the sound speed, and a standard value
for $\alpha =0.01$ is used. The net flux produced by viscous dissipation, $F_{vis}$, is
the only term to balance $F_{rad}$ $-$ unlike D'Alessio \etal (1998) we
have ignored the flux produced by energetic particles ionization.
Then we have the standard relation (see Bell \etal 1997), which holds true for the interior of the
disk where accretion heating occurs:
$$
\sigma T_{vis}^4 = {{3\dot{M} GM_*}\over {8\pi r^3}}\left[1-({R_*\over r})^{1/2}\right] ,
$$
where $\dot{M}$  ~is the mass accretion rate, and $M_*$ and $R_*$ are the stellar mass
and radius.

\section{The Dust}

The properties of the dust affect the wavelength dependence of scattering
and absorption efficiencies.  The temperature in the midplane is sensitive to
the dust scattering albedo (ratio of scattering to total opacity) -- a higher 
albedo would reduce the absorbed stellar flux.
As with our choice of mass accretion rates, we will use dust grains with
properties which best describe the disks of T Tauri stars. 

The modelling of circumstellar disks has always applied dust grain 
properties derived from the interstellar medium. Most commonly used have
been the grain parameters of the Mathis \etal (1977) distribution with
optical constants from Draine \& Lee (1984). However, recent work on
spectral distributions (Whitney \etal 1997) and high-resolution images
(Wood \etal 1998) of T Tauri stars has favored a grain mixture which
Kim, Martin, \& Hendry (1994) derived from detailed fits to the
interstellar extinction law (hereafter KMH). Important grain properties
are the opacity, $\kappa$, the scattering albedo, $\omega$, 
and the scattering asymmetry parameter, $g$.
The latter defines the forward throwing properties of the
dust and ranges from 0 (isotropic scattering) to 1. What sets the KMH grains
apart is that they are more forward throwing ($g$= 0.40($R$), 0.25($K$)), 
and have higher albedo ($\omega$= 0.50($R$), 0.36($K$)) at
each wavelength (optical to near-IR). They are also less polarized, but
that is a property we do not use here. The grain size distribution has the
lower cutoff of KMH (0.005$\mu$m) and a smooth exponential falloff, instead
of an upper cutoff, at 0.25$\mu$m. Since the dust settling time is
proportional to (size)$^{-1}$, we performed calculations with upper
cutoffs of 0.05$\mu$m and 0.1$\mu$m. None of these had any significant
effect on the temperatures.

\section{The Temperature Structure and the Snow Line}
We are interested in planet formation and therefore want to find the
ice condensation line ("snow line") in the midplane of the disk. 
Temperature inversions in the disk's vertical structure (see D'Alessio \etal
1998) may lead to lower temperatures above the midplane, but ice condensation
there is quickly destroyed upon crossing the warmer disk plane.
We define the snow line simply in terms of the local gas-dust temperature
in the midplane, and at a value of 170~K.

In our passive disk, under hydrostatic and radiative equilibrium;
the vertical and radial temperature profiles are similar to those of
Chiang \& Goldreich (1997) and $T(r) \propto r^{-3/7}$. Here is why.
The disk has a concave upper surface (see
Hartmann \& Kenyon 1987) with pressure scale height of the gas at the
midplane temperature, $h$:
$$
{{h}\over r} = \left[{rkT}\over {GM_* \mu m_H}\right] ^{1/2},
$$
where $G$ is the gravitational constant, $\mu$ and $m_H$ are the molecular
weight and hydrogen mass,
$r$ is radius in the disk, and $T$ is the midplane temperature at that
radius. For the inner region (but $r \gg R_*$) of a disk with such concave
shape the stellar
incident flux $F_{irr}(r) \propto \phi(r) \sigma T_*^4 r^{-2}$, where
$\phi(r) \propto r^{2/7}$ (see next section). Here $T_*$ is the effective
temperature of the central star. Then our calculation makes use of
the balance between heating by irradiation and radiative cooling:
$\sigma T^4(r) = F_{irr}(r)$. Therefore our midplane temperature will scale
as $T(r)$= $T_0 r^{-3/7}$~K. This is not surprising, given our standard
treatment of the vertical hydrostatic structure of the disk irradiated
at angles $\phi(r)$. Only the scaling coefficient, $T_0 = 140$, will be
different.
The difference with the Chiang \& Goldreich  model is our treatment 
of the dust
grains $-$ less energy is redistributed inwards in our calculation
and the midplane temperature is lower (Figure 1). The model
with accretion heating is much warmer inwards of 2.5$AU$ where it joins
the no-accretion (passive) model $-$ stellar irradiation dominates.

The result above is for our model with $M_*=0.5M_{\odot}$ and $T_*=4000$~K,
which is standard for T~Tauri stars. It is interesting to see how the snow
line changes for other realistic initial parameters. By retaining the same
dust properties, this can be achieved using scaling relations rather than
complete individual models as shown by Bell \etal (1997) for the pre-main
sequence mass range of 0.5 to 2~$M_{\odot}$. An important assumption at
this point is that we have still retained the same (minimum-mass solar nebula)
disk. With our set of equations,
the midplane temperature coefficient, $T_0$, will be proportional to the stellar mass:
$T_0 \propto M_*^{3/(10-k)}$, where $k$ is a function of the total opacity in
the disk and $0 \leq k < 2$.

On the other hand, different disk masses for a fixed central star 
($M_*=0.5M_{\odot}$ and $T_*=4000$~K) can be modeled for zero accretion
rate, by changing $\Sigma_0$ by a factor of 10 in each direction (defined
in \S 2). Here a remaining assumption is the radial dependence of
$\Sigma \propto r^{-3/2}$; the latter could certainly be $\propto r^{-1}$
(Cameron 1995), or a more complex function of $r$, but it is beyond the
intent of our paper to deal with this. Moreover that we find minimal change
in the midplane temperature in the $r$ range of interest to us (0.1$-$5~AU).
The reason is a near cancellation that occurs between the amount of heating
and increased optical depth to the midplane. One could visualize the vertical
structure of a passive disk for $r = 0.1-5.0$~AU as consisting of three zones:
(1)~optically thin heating and cooling region (dust heated by direct starlight),
(2)~optically thin cooling, but optically thick heating layer, and (3)~the
midplane zone, where both heating and cooling occur in optically thick
conditions. The rate of stellar heating of the disk per unit volume is
directly proportional to the density, and affects the location and temperature
of the irradiation layer. That is nearly cancelled (except for second order
terms) in the mean intensity which reaches the midplane. Therefore we find
that $T(r)$ changes within $\pm 10K$ for a change in disk density ($\Sigma_0$)
of a factor of 10. Note that $T(r)$ is only approximately $\propto r^{-3/7}$
even for $r = 0.1-5.0$~AU; the small effect of density on $T(r)$ has an
$r$ dependence. However, for the purposes of this paper, $i.e.$ our chosen
volume of parameter space, the effect of disk mass on $T(r)$ and the ``snow line" 
is insignificant, and we do not pursue the issue in more detail.
Note that for a disk with a heat source in the midplane, $i.e.$ with an
accretion rate different from zero, the midplane $T(r)$ will be strongly
coupled to the density, roughly $\propto {\rho}^{1/4}$, and will increase
at every $r$ for higher disk masses (e.g. Lin \& Papaloizou 1985).

\section{The Shape of the Upper Surface of the Disk}

The "snow line" calculation in the previous section is made under the
assumption that the upper surface of the disk is perfectly concave and
smooth at all radii, $r$. This is a very good description of such
unperturbed disks, because thermal and gravitational instabilities are
damped very efficiently (D'Alessio \etal 1999).
Obviously this is not going to be the case when an
already formed planet core distorts the disk. But even 
a small distortion of the disk's surface may affect the
thermal balance. The distortion need only be large enough compared to
the grazing angle at which the starlight strikes the disk, $\phi(r)$:
$$
\phi(r) = {{0.4R_*}\over {r}}+r{{d}\over {dr}}({{h}\over {r}}),
$$
where $h$ is the local scale height.
This small angle has a minimum
at 0.4$AU$ and increases significantly only at very large distances:
$\phi(r) \approx 0.005r^{-1} + 0.05r^{2/7}$ (e.g. see Chiang \& Goldreich
1997).

The amount of compression due to the additional mass of the planet, $M_p$, 
will depend on the Hill radius, $R_H = r ({{M_p}\over {M_*}})^{1/3}$,
and how it compares to the local scale height, $h$. The depth of the depression
will be proportional to $(R_H/h)^3$. The resulting depressions
(one on each side) will be in the shadow from the central star, with a
shade area dependent on the grazing angle, $\phi(r)$. The solid angle
subtended by this shade area from the midplane determines the amount of cooling
and the new temperature in the sphere of influence of the planet core.
The question then arises,
if during the timescale preceeding the opening of the gap the midplane
temperature in the vicinity of the accreting planet core could drop
below the ice condensation limit even for orbits with $r$ much shorter
than the "snow line" radius in the undisturbed disk. 
The answer appears to be affirmative and a runaway develops whereby local
ice condensation leads to rapid growth of the initial rocky core, which
in turn deepens the depression in the disk and facilitates more ice
condensation inside the planet's sphere of influence.
Details about the
instability which develops in this case will be given in a separate paper.

\section{Conclusion}
When the large fraction of close-in extrasolar giant planets became
apparent, we thought of questioning the standard notion of a distant "snow
line" beyond 3$AU$ in a protoplanetary disk. Thence comes this paper.
We revisited the issue by paying attention to the stellar irradiation
and its radiative effects on the disk, thus limiting ourselves to
passive or low accretion rate disks.

We find a snow line as close as 0.7$AU$ in a passive disk, and not
much further away than 1.3$AU$ in a disk with 10$^{-8}M_{\odot} {\rm yr}^{-1}$
accretion rate for $M_*=0.5M_{\odot}$. The result is robust regardless
of different reasonable model assumptions $-$ similar values
could in principle be inferred from existing disk
models (Chiang \& Goldreich 1997; D'Alessio \etal 1998). For more massive
(and luminous) central stars, the snow line shifts outwards: to 1.0$AU$ 
(1$M_{\odot}$) and 1.6$AU$ (2$M_{\odot}$). The effect of different disk
mass is much smaller for passive disks $-$ the snow line shifts inwards
by 0.08$AU$ for ${\Sigma}_0$= 10$^4$g~cm$^{-2}$. Our results
differ from existing calculations (in that they bring the
snow line even closer in), because the dust grains properties we
used have higher albedo and more forward throwing. The dust grains
and the disk models we used are typical of T Tauri stars of age $\sim$1~Myr.
So our conclusion is, that if such T Tauri disks are typical of
protoplanetry disks, then the snow line in them could be as close-in
as 0.7 AU.

Our estimate of the snow line is accurate to within 10\%, once the model
assumptions are made. These assumptions are by no means good or obvious,
and can change the numbers considerably. For a passive disk model, the assumptions
that need to be justified are: the equilibrium of the disk, the lack of
dust settling (i.e. gas and dust are well mixed), the used KMH properties 
of the dust grains, and the choice of molecular opacities. For a low
accreting disk model, one has to add to the above list: the choice of
viscous dissipation model (and $\alpha$=0.01).

Finally, we note that these estimates reflect a steady-state disk in
hydrostatic equilibrium. The disk will get disturbed as planet formation
commences, which may affect the thermal balance locally given the small
value of the grazing angle, $\phi$. For a certain planet core mass,
an instability can develop at orbits smaller than 1 AU which can lead 
to the formation of giant planets in situ. What is then the determining
factor for the division between terrestrial and giant planets in our
Solar System remains unexplained (as it did even with a snow line at 2.7$AU$).

\acknowledgements{
We thank N. Calvet, B. Noyes, S. Seager, and K. Wood for reading the
manuscript and helpful discussions, and the referee for very thoughtful
questions.
}

\begin{figure}[!t]
\plotfiddle{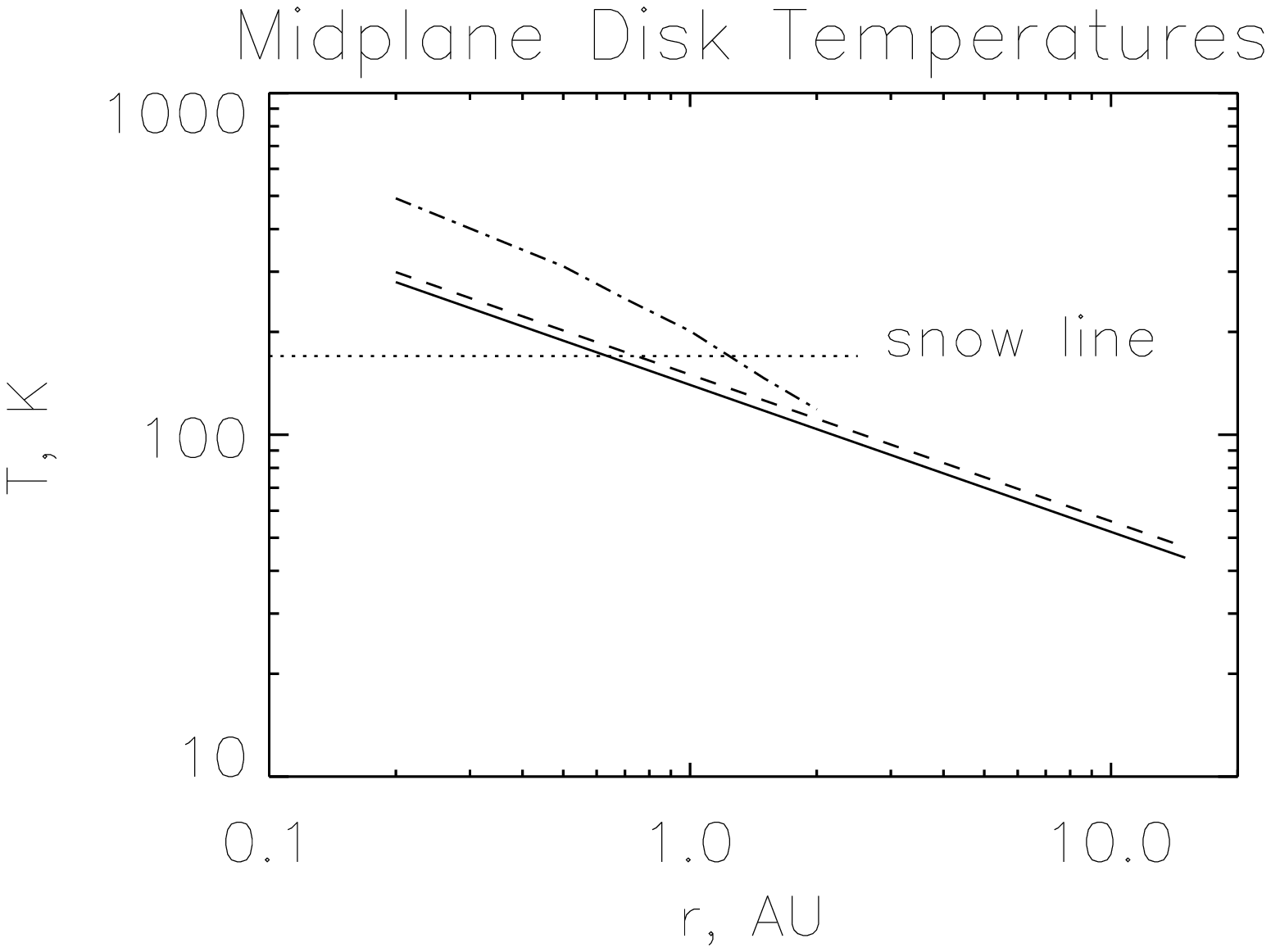}{7.5cm}{0}{100}{100}{-305}{-370}
\caption{
Temperature profiles in the midplane for the two models
computed - passive (solid) and with accretion heating (dash-dot).
The passive disk model of Chiang \& Goldreich 1997 is shown also
(dashes).}
\end{figure}

\end{document}